\documentclass[twocolumn, secnumarabic, amssymb, amsmath,nobibnotes, aps, prl, showpacs]{revtex4}

\begin{document}

\title{Thermodynamics of Means} 
\author{B. H. Lavenda}
\email{bernard.lavenda@unicam.it}
\affiliation{Universit\'a degli Studi, Camerino 62032 (MC) Italy}
\date{\today}
\newcommand{\sumi}{\sum_{i=1}^{n}\,}
\newcommand{\sumj}{\sum_{j=1}^{n}\,}
\newcommand{\sumk}{\sum_{i=1}^{k}\,}
\newcommand{\sumgr}{\sum_{i=1}^{>}\,}
\newcommand{\sumle}{\sum_{i=1}^{<}\,}
\newcommand{\half}{\mbox{\small{$\frac{1}{2}$}}}
\newcommand{\third}{\mbox{\small{$\frac{1}{3}$}}}\newcommand{\twothirds}{\mbox{\small{$\frac{2}{3}$}}}
\begin{abstract}
 Thermodynamics of power means applies to an ideal quantum gas which may be nonextensive.  Transition to an ideal classical gas occurs when the empirical temperature exponents of the internal energy and absolute temperature coalesce. Limiting processes are pure heat conduction and pure deformations. Largest and smallest mean final volumes occur for isothermal and adiabatic processes, respectively. The increment in the heat admits two integrating factors which yield conserved quantities for adiabatic processes. Energy-conserving equilibrations yield the largest final means possible, while the second law follows from the property that the power means are monotonically increasing functions of their order.  In the ideal classical gas limit, the change in the average entropy is  proportional to the difference between the  Shannon and R\'enyi entropies for nonextensive, isothermal systems that are multifractal in nature.
\end{abstract}
\maketitle
Thermodynamics developed largely independently of all the classical branches of mathematics. Unlike other laws of physics, the second law is an inequality, whose generality rests on there not being macroscopic violations of that inequality. Inequalities is a well-established branch of mathematics \cite{HLP}, and it would be of interest to show that the laws of thermodynamics can be derived from well-known inequalities.\par Carath\'eodory's formulation of the second law, in its most general nonstatical form \cite{Chandrasekhar}, states that in every neighborhood of any given state, there exists adiabatically inaccessible states even when nonstatical processes are permitted. If we envision an entropy change, $\Delta S$, in which $dQ=0$ (adiabatic), and at constant volume, the entropy of the system can only increase or decrease. Since this is true no matter what the initial state is, it must be always increase or always decrease. In order to determine which of the two possibilities is the correct one, recourse must be made to experiment---albeit a single experiment suffices. \par
It would be more gratifying if the entropy change could be determined by a mathematical inequality, that would be independent of any particular process or type of substance considered. In a little known paper, Cashwell and Everett \cite{Cashwell} do just this without exploiting the full impact of their findings. What they do is to show that the final mean temperature of order-$q$ in an energy-conserving equilibration is necessarily greater than the mean of order-$(q-r)$, where $q>r$, of an entropy-conserving equilibration. Thus, the increase in entropy is due to the property that power means are increasing functions of their order \cite{Burrows}.\par
 \par
 In a series of papers, Landsberg \cite{PTL78}, and Landsberg and Pe\v{c}ari\v{c} \cite{PTL-bis}, rediscover the Cashwell-Everett results in the case of constant, but not necessarily positive, heat capacities. Both Cashwell-Everett and Landsberg were anteceded  by Sommerfeld \cite{Sommerfeld} who derived the arithmetic-geometric mean inequality for the temperature from the second law as an exercise.  Sidhu \cite{Sidhu} generalized Landsberg's result to arbitrary power means of order-$q>1$, or, equivalently, to systems whose heat capacities are some positive power of the temperature. Negative orders were excluded on the basis that it would contradict the third law of thermodynamics. \par
From a mathematical viewpoint, there is no qualitative distinction between the arithmetic mean, $q=1$, and power means, $q>1$. However, there is a thermodynamic distinction insofar as the former applies to an ideal \emph{classical} gas (ICG), while the later to an ideal \emph{quantum} gas (IQG). \par
Our starting point will be the purely mechanical equation of state \cite{Einbinder,PTL61}
\begin{equation}
pV=\frac{r}{q-r}U, \label{eq:pV}
\end{equation}
which we will assume to hold for IQG, although a formally analogous relation will appear for ICG, where $p$ is the pressure, $V$ the volume of the container, and $U$ the internal energy. $r$ and $q$ are positive rational numbers, where $q>r$. \par
Differentiating (\ref{eq:pV}) with respect to $V$, at a constant empirical temperature $t$, and using a combination of the Gibbs equation and  a Maxwell relation result in
\begin{eqnarray*}
V\left(\frac{\partial p}{\partial V}\right)_t+p & = &
\frac{r}{q-r}\left(\frac{\partial U}{\partial V}\right)_t\\
& = & \frac{1}{q-r}\left\{t\left(\frac{\partial p}{\partial t}\right)_V-rp\right\}.
\end{eqnarray*}
This partial differential equation has the general solution
\[
p=\frac{r}{q-r}V^{-q/(q-r)}Z(z), \]
or, equivalently,
\begin{equation}
U=V^{-r/(q-r)}Z(z), \label{eq:U-bis}
\end{equation}
on the strength of (\ref{eq:pV}), where $z=tV^{1/(q-r)}$.\par At this point we can impose the property of homogeneity of the internal energy by requiring
\begin{equation}
\left(\frac{\partial p}{\partial V}\right)_t=0, \label{eq:Clap}
\end{equation}
which, in view of (\ref{eq:pV}), implies that $U$ is first-order homogeneous. It will be appreciated that (\ref{eq:Clap}) is a condition for phase equilibrium. Varying the volume at constant temperature leaves the vapor pressure constant by having the liquid evaporate or the vapor condense. This fine balance  keeps the pressure constant and preserves the homogeneity of the internal energy. Moreover, it determines $Z= z^q$, where we take the constant of proportionality equal to unity for simplicity sake. The condition of first-order homogeneity is, however,  not necessary.\par
From (\ref{eq:U-bis}) we can write
\begin{equation}
UV^{r/(q-r)}= z^{\alpha}= t^{\alpha}V^{\alpha/(q-r)}=:Z(z), \label{eq:Z}
\end{equation}
by assuming a power expression for $Z(z)$.  The particular choice, $q=\alpha$, renders the internal energy homogeneous, and the system extensive.\par
Solving for $U$,
\begin{equation}
U= t^{\alpha}V^{(\alpha-r)/(q-r)}, \label{eq:U}
\end{equation}
and using (\ref{eq:pV}) give
\begin{equation}
p=\frac{r}{q-r}t^{\alpha}V^{(\alpha-q)/(q-r)}. \label{eq:p}
\end{equation}
The analogy with the Joule-Thomson effect will help clarify the role of the exponents.  The condition that the entropy be an exact differential is
\begin{equation}
\left(\frac{\partial H}{\partial p}\right)_t=V-\frac{t}{r}\left(\frac{\partial V}{\partial t}\right)_p=:C_p\mu, \label{eq:JT}
\end{equation}
where $H$ is the enthalphy [cf. (\ref{eq:H}) below], and $C_p$ is the heat capacity at constant pressure. Using (\ref{eq:p}), we evaluate the Joule-Thomson coefficient, $\mu$, as
\begin{equation}
\mu C_p=\frac{q}{r}\left(\frac{\alpha-r}{\alpha-q}\right)V,\label{eq:mu}
\end{equation}
since
\begin{equation}
\left(\frac{\partial V}{\partial t}\right)_p=-\frac{V}{t}\left(\frac{q-r}{\alpha-q}\right). \label{eq:V/t}
\end{equation}
When $\alpha>q$, $\mu$ is positive and there is a cooling effect. Alternatively,
when $q>\alpha>r$, $\mu$ is negative and there is a heating effect. When the system is extensive, $\alpha=q$, (\ref{eq:mu}) implies that $C_p=\infty$. The inversion point occurs when $\alpha=r$. For an extensive system, (\ref{eq:Clap}) shows that the pressure depends only on the temperature, and is independent of the volume. Since constant pressure implies constant temperature, when heat is added to the system  the temperature must remain fixed, and only the volume can change. This implies $C_p=\infty$ for all extensive IQG.
\par
When the internal energy (\ref{eq:U}) is not a \emph{linear\/} function of the empirical temperature \emph{alone\/}, the gas and absolute temperature scales do not coincide. For an ICG, the constant of proportionality between the internal energy and the empirical temperature is some multiple of the number of particles, $N$, which is a constant. Rather, for an IQG, the number of particles is variable, being a function of the  temperature.\par
The only thing that the zeroth law demands is that the empirical temperature be the same when two systems arrive in a state of mutual thermal equilibrium. Once the empirical scale has been chosen, the absolute temperature must be a monotonically increasing function of it, say $T(t)=t^r$ for $r>0$.\par
In order to render the increment in the heat,
\begin{eqnarray}
dQ & = & dU+p\,dV\label{eq:dQ} \\
 & = & \alpha t^rz^{\alpha-r}d\log z=\alpha V^{-r/(q-r)}z^{\alpha}d\log z\nonumber,
\end{eqnarray}
a perfect differential, we may either divide through by $t^r$ to obtain
\begin{equation}
\frac{\alpha}{\alpha-r}d\left(t^{\alpha-r}V^{(\alpha-r)/(q-r)}
\right)
=\frac{dQ}{t^r}=:dS,
\label{eq:dS}
\end{equation}
or multiply  through by $V^{r/(q-r)}$ to get
\begin{equation}
d\left(t^{\alpha}V^{\alpha/(q-r)}\right)= V^{r/(q-r)} dQ=:dZ. \label{eq:dZ}
\end{equation}
\par
The potential  (\ref{eq:Z}) is a higher power of $z$ than the entropy,
\begin{equation}
S=\frac{\alpha}{\alpha-r} z^{\alpha-r}=\frac{\alpha}{\alpha-r} t^{\alpha-r}V^{(\alpha-r)/(q-r)}, \label{eq:S}
\end{equation}
which becomes extensive when $\alpha=q$. Since
\begin{equation}
H=U+pV=\frac{q}{q-r}t^{\alpha}V^{(\alpha-r)/(q-r)}, \label{eq:H}
\end{equation}
the difference between (\ref{eq:H}) and $t^r$ times (\ref{eq:S}), 
\begin{equation}
G=U+pV-t^rS=\left(\frac{q}{r}-\frac{\alpha}{r}\frac{q-r}{\alpha-r}\right)pV, \label{eq:G}
\end{equation}
shows that the Gibbs free energy, $G$, 
is a measure of the degree of nonextensivity. It is well known that (\ref{eq:V/t}) diverges when  the chemical potential is identically zero \cite[p. 229]{PTL61}. This occurs for an extensive IQG.\par
For a nonrelativistic IQG, irrespective of whether it obeys Bose or Fermi statistics, $\alpha=q$,  $q/r=\mbox{\small{$\frac{5}{2}$}}$, and we shall set $r=1$ in order to make the comparison with conventional formulas as close as possible. Apart from constant factors, $U=T^{5/2}V$, and $Z=V^{2/3}U=T^{5/2}V^{5/3}$. Since the latter must be a power of $z=TV^{2/3}$ only, $U$ cannot contain any arbitrary constants. Conventionally, such an arbitrary constant is identified with the chemical potential at absolute zero. Since conventional IQG are extensive, the chemical potential vanishes identically on account of (\ref{eq:G}).\par
In an adiabatic process, the entropy $S=\mbox{\small{$\frac{5}{3}$}}z^{3/2}=\mbox{\small{$\frac{5}{3}$}}T^{3/2}V$, and the potential, $Z=z^{5/2}=\mbox{\small{$\frac{3}{2}$}}pV^{5/3}$, are constant. The pressure, $p=\mbox{\small{$\frac{2}{3}$}}T^{5/2}$, is independent of the volume, and satisfies the homogeneity condition, (\ref{eq:Clap}).
\par
A new adiabatically conserved potential, (\ref{eq:Z}), has appeared,  for which the cell size, $R=V^{1/(q-r)}$, raised to the power $r$, is the integrating factor for the heat,
\[
R^rdQ=dZ, \]
which puts it on par with that of the second law, (\ref{eq:dS}). Since both (\ref{eq:Z}) and (\ref{eq:S}) are constant for adiabatic processes, they provide no criterion for the accessibility of states under adiabatic transitions. For processes of pure heat conduction,  $U$- and $Z$-conservation equilibrations give the greatest mean temperature possible, and the evolution is determined by the second law, because the power means are monotonic increasing functions of their order. However, for processes involving pure deformations, $U$- and $Z$-conservation equilibriations will not be equivalent.
\par 
Consider a system comprised of $n$ cells, adiabiatically isolated from the environment. Initially the walls of the cells are rigid and adiabatic. When the walls are replaced by deformable, diathermal ones, there will be  probabilities, $p_i$ that the cells will have linear dimensions, $R_i$ and temperatures, $t_i$. Probabilities enter naturally when dealing with processes of heat exchange:  Heat  is the uncontrollable form of work \cite{JJ}, and temperature is its measure.\par
Carath\'eodory's principle says that states arbitrarily near to one another may be adiabatically inaccessible; it does not say what those states are. This must come from an additional assumption that only those states that are adiabatically  accessible from a given state are those for which the internal energy increases \cite{Buchdahl},
\begin{eqnarray}
\overline{\Delta U} & = & -\frac{r}{q-r} z^{\alpha}\sumi p_i\int_{V_i}^{V_f}V^{-q/(q-r)}dV\nonumber\\
& = & z^{\alpha}\left\{V_f^{-r/(q-r)}-\sumi p_iV_i^{-r/(q-r)}
\right\}\nonumber\\
& =& \mbox{const.}\times\left\{t^r_f-\sumi p_it_i^r\right\}\ge0,
\label{eq:DeltaU}
\end{eqnarray}
by the work that must be done. The last line follows from the condition of adiabaticity, $z=tV^{1/(q-r)}=\mbox{const}$. Inequality (\ref{eq:DeltaU}) asserts that adiabatic transitions to neighboring states are possible so long as the final temperature is greater than the mean temperature, $t(r)=\left(\sumi p_it_i^r\right)^{1/r}$.\par This is tantamount to requiring the final mean volume to be greater than $\left(\sumi p_iV_i^{-r/(q-r)}\right)^{-(q-r)/r}$. In an adiabatic, energy-conserving equilibration, the final mean volume,
\[
V_f=\left(\sumi p_iV_i^{-r/(q-r)}\right)^{-(q-r)/r}\rightarrow V_{\min},
\]
in the limit as $q\rightarrow r$, where $V_{\min}$ is the smallest cell volume. The final mean volume, $V_f$, is a decreasing function of its order. On account of the adiabatic constraint, the mean temperature is  the highest mean temperature, $t(q)$. The smaller the difference, $q-r$, the more weight is given to  smaller cell volumes. 
\par
Alternatively, if no work is performed, the average change in the internal energy cannot increase,
\begin{eqnarray}
\overline{\Delta U} & = & V^{\alpha-r)/(q-r)}\sumi p_i\int_{t_i}^{t_f}dt^{\alpha}\nonumber\\
& = & V^{(\alpha-r)/(q-r)}\left\{t_f^{\alpha}
-\sumi p_it_i^{\alpha}\right\}\le0, \label{eq:DeltaU-bis}
\end{eqnarray}
in a process involving only pure thermal conduction. This says that the final common mean temperature of the cells cannot be greater than the mean temperature, $t(\alpha)$, which is the greatest mean temperature in nonextensive systems with $\alpha>q$. Inequality (\ref{eq:DeltaU-bis}) can be related to Jensen's inequality, and constitutes a criterion for spontaneous transition \cite{BHL}. Consequently, $\overline{\Delta U}=0$ coincides with $\overline{\Delta Z}=0$ for  processes in which no work is done, where the highest final mean temperature is reached.
\par
Contrarily, the largest mean volume coincides with the largest cell volume, $V_{\max}$, in an isothermal energy-conserving equilibration in the same limit as $q\downarrow r$. The energy equilibriating conservation condition for an isothermal process gives the final mean volume \begin{equation}V_f=\left(\sumi p_iV_i^{(\alpha-r)/(q-r)}\right)^{(q-r)/(\alpha-r)}\rightarrow V_{\max}
\label{eq:Vmax}
\end{equation}
in the limit as $q\rightarrow r$ if $\alpha>q$. The final volume is now an increasing function of its order. The larger the order, the more weight is given to larger cell volumes $V_i$.\par 
The final mean volume in a $Z$-conserving equilibration,
\[V_f=\left(\sumi p_iV_i^{\alpha/(q-r)}\right)^{(q-r)/\alpha},\]
will be greater than in a $U$-conserving equilibration, (\ref{eq:Vmax}), but will tend to the same value, $V_{\max}$, in the limit as $q\downarrow r$.
\par
The limit  $q\downarrow r$ will be shown to be the transition from an IQG to an ICG in nonextensive systems with $\alpha>q$, or extensive systems in general. In that limit, an adiabatic transition corresponds to the smallest volume, while an isothermal transition corresponds to the greatest final volume. Hence, adiabatic to isothermal changes cover the entire spectrum of mean values of the volume, analogous to a polytrope, where adiabats and isotherms correspond to zero and infinite heat capacities, respectively.
\par
If the temperatures and sizes of the cells are similarly ordered, so that cells of higher temperatures have larger sizes, in compliance with Charles' law, 
\begin{eqnarray}
z(q) & = & \left(\sumi p_it_i^qR_i^q\right)^{1/q}>t(q)R(q)\nonumber\\
& = &
\left(\sumi p_it_i^q\right)^{1/q}\left(\sumi p_iR_i^q\right)^{1/q}, \label{eq:Cheby}
\end{eqnarray}
unless all the temperatures, $t_i$, and cell sizes, $R_i$, are equal. If there is some constraint that makes the temperatures decreasing functions of the cell sizes, then the discrete analog of \v{C}hebyshev's inequality, (\ref{eq:Cheby}), is reversed \cite[pp. 43--44]{HLP}.\par
It therefore follows that the terms in the hierarchy
\[z_h>z(q)>z(q-r)>z(r)>z_c,\]
are greater than the corresponding terms in the hierarchy
\[t_hR_{\max}>t(q)R(q)>t(q-r)R(q-r)>t(r)R(r)>t_cR_{\min}.\]
This is implied by the hierarchy
\[t_h>t(q)>t(q-r)>t(r)>t_c,\]
in the case of pure thermal conduction, or, in the case of an isothermal process by
\[R_{\max}>R(q)>R(q-r)>R(r)>R_{\min},\]
where $R_{\max}$ and $R_{\min}$ are the largest and smallest cell lengths, respectively.\par
In a $Z$-equilibrating conservation, 
\begin{equation}
\overline{\Delta Z}=\sumi p_i\int_{z_i}^{z_f}dZ(z)=
z^{\alpha}_f-\sumi p_iz_i^{\alpha}=0, \label{eq:I-law}
\end{equation}
imposing $z_f=z(\alpha)$. The fact that power means are monotonically increasing functions of their order,
\[
\left(\sumi p_iz_i^{\alpha}\right)^{1/\alpha}>
\left(\sumi p_iz_i^{\alpha-r}\right)^{1/(\alpha-r)},
\]
for $\alpha>r$, guarantees the second law
\begin{eqnarray}
\overline{\Delta S} & = & \sumi p_i\int_{z_i}^{z_f}
\frac{dZ(z)}{z^r}\nonumber\\
& = & \frac{\alpha}{\alpha-r}\left\{z^{\alpha-r}(\alpha)-
\sumi p_iz_i^{\alpha-r}\right\}\ge0. \label{eq:II-law}
\end{eqnarray}
This can be considered the essence of the second law for power means \cite{Cashwell}.\par
The transition between IQG and ICG is usually considered to take place in the high temperature limit. Here, we show that the transition takes place in the limit as $\alpha\downarrow r$. In this limit (\ref{eq:II-law}) becomes indeterminate, of the form $0/0$. Invoking 
l'H\^opital's rule, we get
\begin{equation}
\overline{\Delta S}=\log\left(\sumi p_iz_i^{r}\right)-r\sumi p_i\log z_i. \label{eq:Hopital}
\end{equation}
It is important to observe that the limit $q\downarrow r$ cannot be taken since it would leave the cell sizes $R_i$ undefined.\par Rather, if we consider an extensive system, where $\alpha=q$, and consider only isochoric processes, then (\ref{eq:II-law}) becomes
\[
\overline{\Delta s}=\frac{q}{q-r}\left\{\left(\sumi p_it_i^q\right)^{(q-r)/q}-\sumi p_it_i^{q-r}\right\}, \]
upon dividing through by the constant volume, $V$, where $s$ is the entropy density. Now, taking the limit as $q\downarrow r$, and employing l'H\^opital's rule we get
\[
\lim_{q\downarrow r}\overline{\Delta s}=r\log\left[\frac{t(r)}{t(0)}\right]=\log\left[\frac{u(1)}{u(0)}\right]>0.\]
 The quantities $u(1)=\sumi p_iu_i$, and $u(0)=\prod_{i=1}^n u_i^{p_i}$ are now to be interpreted as  \emph{molar\/} energies, and not as  \emph{densities\/} \cite{BHL91}. The second law inequality now follows  from the arithmetic-geometric mean inequality \cite{Sommerfeld}, implying that the maximum amount of work has been done in going from the initial state, characterized by the mean energy, $u(1)$, to the final state, with energy $u(0)$.\par
The ICG limit also allows a connection to be made with multifractals and information theory \cite{BHL98}, in the isothermal processes occurring in nonextensive systems. In this case, (\ref{eq:Hopital}) becomes
\begin{equation}
\overline{\Delta S}=\log\left(\sumi p_iR_i^{D\tau}\right)-D\tau\sumi p_i\log R_i, \label{eq:II-bis}
\end{equation}
where we have set the exponent, $r=D\tau$. The exponent $D$ is the Hausdorff dimension, defined as
\begin{equation}
\sumi R_i^D=1. \label{eq:Haus}
\end{equation}
Condition (\ref{eq:Haus}) plays a role analogous to the Kraft (in)equality for a uniquely decipherable code.
\par
If we consider the exponent, $\tau<1$, then we can apply H\"older's inequality in the reverse form \cite[p.25]{HLP}
\begin{eqnarray*}
\left[\sumi\left(p^{1/\tau}_iR_i^D\right)^{\tau}\right]^{1/\tau}
\left[\sumi p_i^{-(1/\tau)\cdot\tau/(\tau-1)}\right]^{(\tau-1)/\tau}\\
  \le 
\sumi R_i^D=1,
\end{eqnarray*}
on the strength of (\ref{eq:Haus}). Setting $\tau=(\alpha-1)/\alpha<1$, with $\alpha>1$, H\"older's inequality becomes
\begin{equation}\sumi p_iR_i^{D\tau}\le\left(\sumi p_i^{\alpha}\right)^{1/\alpha}.\label{eq:Holder}
\end{equation}
\par
It is easy to see that we have equality in (\ref{eq:Holder}) if
\[R_i^D=\frac{p_i^{\alpha}}{\sumi p_i^{\alpha}},\]
or
\begin{equation}
D\log R_i=\alpha\log p_i-\log\sumi p_i^{\alpha}, \label{eq:optimal}
\end{equation}
which also satisfies the definition of the Hausdorff dimension, (\ref{eq:Haus}). Multiplying (\ref{eq:optimal}) through by $p_i$, summing, and introducing the result into (\ref{eq:II-bis}) give
\begin{equation}
\overline{\Delta S}=(\alpha-1)\left(S_1-S_{\alpha}\right)>0, \label{eq:S-info}
\end{equation}
where $S_1$ and $S_\alpha$ are the Shannon,
\[S_1=-\sumi p_i\log p_i,\]
and R\'enyi,
\[S_\alpha=\frac{1}{1-\alpha}\log\sumi p_i^{\alpha},\]
entropies of order $1$ and $\alpha$, respectively. The inequality in (\ref{eq:S-info}) is due to the fact that for $\alpha>1$ the Shannon entropy is greater than the R\'enyi entropy, while, for $\alpha<1$, the converse is true. Hence, the second law (\ref{eq:S-info}) is always satisfied. Moreover, in the limit as $\alpha\rightarrow1$, l'H\^opital rule shows that
\[S_1=\lim_{\alpha\rightarrow1}S_{\alpha}=-\sumi p_i\log p_i.\]
In this limit, the average entropy difference, (\ref{eq:S-info}), vanishes.\par Hence, (\ref{eq:S-info}) shows that in the limit of an isothermal ICG, the average  entropy difference is always proportional to the absolute value of the difference between the Shannon and R\'enyi entropies, when $D$ is identified as the Hausdorff dimension.\par The generalization of the Hausdorff dimension to multifractals, where the generator of cell sizes of lengths $R_i$ with probabilities, $p_i$, require \emph{two} exponents \cite{Hal}, 
\begin{equation}\sumi p_i^{\alpha}R_i^{D_{\alpha}(1-\alpha)}=1,\label{eq:Haus-bis}
\end{equation}
where $D_{\alpha}$ is supposed to be some generalization of the Hausdorff dimension, $D$.\par
If $\{p_i\}$ is a complete distribution, and $\alpha$ is restricted to the open interval $(0,1)$ in order ensure that the R\'enyi entropy be concave, then  the usual H\"older inequality, and condition (\ref{eq:Haus-bis}), give
\begin{equation}\sumi R_i^{D_{\alpha}}\ge1.\label{eq:Holder-bis}
\end{equation}
Since for $\alpha=1$, (\ref{eq:optimal}) becomes
\[D_1=\frac{S_1}{\sumi p_i\log(1/R_i)},\]
 it was thought \cite{Hal} that $D_\alpha$ should be related to the R\'enyi entropy in a similar form, viz.,
\[D_{\alpha}=\frac{S_{\alpha}}{\sumi p_i\log(1/R_i)}.\]
\par
This can be obtained by averaging
\begin{equation}
D_{\alpha}(1-\alpha)\log R_i=-\log\sumi p_i^{\alpha}.\label{eq:D}
\end{equation}
Exponentiating both sides, multiplying by $p_i^{\alpha}$, and summing  does give (\ref{eq:Haus-bis}). But, since the right side of (\ref{eq:D}) is independent of the index $i$, so too must be the left side. This means that all the cell sizes have the same length
\[R^{D_{\alpha}}=\left(\sumi p_i^{\alpha}\right)^{-1/(1-\alpha)}=e^{-S_{\alpha}}.\]
In view of condition (\ref{eq:Holder-bis}), this would imply
\[S_0\ge S_{\alpha},\]
showing that equal probabilities maximize the entropy: The greatest entropy is the Hartley entropy, $S_0=\log n$, which is frequency independent. \par

\end{document}